\documentclass[preprint,prd]{revtex4-1}
\usepackage{epsfig}
\usepackage{graphicx}
\usepackage{dcolumn}
\usepackage{amsmath}
\usepackage{latexsym}
\usepackage{amssymb,amsfonts,verbatim}
\usepackage{hyperref}

\def\bfl{\begin{flushleft}}
\def\efl{\end{flushleft}}
\def\bfr{\begin{flushright}}
\def\efr{\end{flushright}}
\def\bc{\begin{center}}
\def\ec{\end{center}}
\def\be{\begin{equation}}
\def\ee{\end{equation}}
\def\ba{\begin{eqnarray}}
\def\ea{\end{eqnarray}}
\def\baa#1{\begin{array}{#1}}
\def\eaa{\end{array}}
\def\bw{\begin{widetext}}
\def\ew{\end{widetext}}

\newcommand\beq{\begin{equation}}      
\newcommand\beqnn{\begin{eqnarray*}}   
\newcommand\beqa{\begin{eqnarray}}     
\newcommand\beqann{\begin{eqnarray*}}  

\newcommand\eeq{\end{equation}}        
\newcommand\eeqnn{\end{eqnarray*}}     
\newcommand\eeqa{\end{eqnarray}}       
\newcommand\eeqann{\end{eqnarray*}}    

\newcommand{\diffpa}[1]{\frac{\partial}{\partial {#1}}}                                

\newcommand{\ket}[1]{\left| #1 \right\rangle}                                               
\newcommand{\bra}[1]{\left\langle #1\right|}                                               
\newcommand{\ketr}[1]{\left| #1 \right)}                                                    
\newcommand{\brar}[1]{\left( #1\right|}                                                    
\newcommand{\overlap}[2]{\left\langle {#1} | {#2} \right\rangle}                            
\newcommand{\matrel}[3]{\left\langle {#1} \right| {#2} \left| {#3}\right\rangle}            
\newcommand{\overlapr}[2]{\left( {#1} | {#2} \right)}                                       
\newcommand{\matrelr}[3]{\left( {#1} \right| {#2} \left| {#3}\right)}                       

\def\bd {b^{\dagger}}
\def\bl {B_{L}}
\def\bld {B_{L}^{\ddagger}}
\def\br {B_{R}}
\def\brd {B_{R}^{\ddagger}}
\def\zb {\bar{z}}
\def\vb {\bar{v}}

\newcommand{\tr}[1]{\textnormal{tr}_{c} \left( #1 \right) }                                 
\newcommand{\func}[2]{\textnormal{#1}\left( #2 \right)}                                     
\newcommand{\modsq}[1]{\left| #1 \right|^{2}}
\def\nl {\nonumber \\}

\let\oldsqrt\sqrt
\def\sqrt{\mathpalette\DHLhksqrt}
\def\DHLhksqrt#1#2{%
\setbox0=\hbox{$#1\oldsqrt{#2\,}$}\dimen0=\ht0
\advance\dimen0-0.2\ht0
\setbox2=\hbox{\vrule height\ht0 depth -\dimen0}%
{\box0\lower0.4pt\box2}}

\begin{document}

\begin{flushright}
\end{flushright}

\title{Noncommutative quantum mechanics -- a perspective on structure and spatial extent}
\date{\today}
\author{C M Rohwer$^{b}$, K G Zloshchastiev$^{a}$, L Gouba$^{a}$ and \footnote{Corresponding author: fgs@sun.ac.za}F G Scholtz$^{a,b}$}
\affiliation{$^a$National Institute for Theoretical Physics (NITheP),
Stellenbosch 7600, South Africa\\
$^b$Institute of Theoretical Physics,
University of Stellenbosch, Stellenbosch 7600, South Africa}

\begin{abstract}
\noindent
We explore the notion of spatial extent and structure, already alluded to in earlier literature, within the formulation of quantum mechanics on the noncommutative plane.  Introducing the notion of average position and its measurement, we find two equivalent pictures: a constrained local description in position containing additional degrees of freedom, and an unconstrained nonlocal description in terms of the position without any other degrees of freedom.   Both these descriptions have a corresponding classical theory which shows that the concept of extended, structured objects emerges quite naturally and unavoidably there. It is explicitly demonstrated that the conserved energy and angular momentum contain corrections to those of a point particle.  We argue that these notions also extend naturally to the quantum level. The local description is found to be the most convenient as it manifestly displays additional information about structure of quantum states that is more subtly encoded in the nonlocal, unconstrained description. Subsequently we use this picture to discuss the free particle and harmonic oscillator as examples.
\end{abstract}

\maketitle


\section{Introduction and background}
\label{intro}

The continuing search for a consistent theory of quantum gravity has identified a number of generic features that such a theory must have.  It has become clear that the notion of point-like local interactions has to be replaced by nonlocal interactions and that our notion of space-time beyond the Planck scale probably needs drastic revision.  In the context of string theory nonlocal interactions are introduced by means of spatially extended objects, be it strings or branes \cite{ref1}.  Another setting in which non local interactions occur naturally is that of noncommutative quantum field theories, which adopts the point of view that the structure of space-time beyond the Planck scale has a noncommutative nature \cite{dop}. One may quite naturally then pose the question whether noncommutative quantum field theories admit a description in terms of extended objects.  Indeed, for a free particle moving in the noncommutative plane the notion of physical extent and structure was already made explicit in \cite{suss}, where it was shown that this system can be thought of as two oppositely charged particles interacting through a harmonic potential and moving in a strong magnetic field. This construction therefore suggests that noncommutative quantum mechanics may in some way encode the notion of physical extent and structure and that it may also admit a description in terms of extended objects.  With this in mind we explore here the notion of extended objects in noncommutative quantum mechanics in the general formalism of \cite{laure}, and argue that the description of noncommutative quantum mechanics in terms of extended, structured objects is indeed a natural one. The essence of our approach is to introduce an average position for the particle, which can be done very precisely and unambiguously in the formalism of \cite{laure}.  We find that there are two equivalent scenarios in which this notion of extendedness can be realized, the one an unconstrained but non local description in the position, in the sense that it requires knowledge of the wave function and all its derivatives at a given point, and the other a manifestly local, but constrained description invoking additional degrees of freedom.

We begin by recalling the basic construction of a noncommutative quantum system as outlined in \cite{laure}. Restricting to two dimensions, the coordinates of noncommutative configuration space satisfy the commutation relation
\begin{equation}
[\hat{x}, \hat{y}] = i\theta,
\label{xycommutator}
\end{equation}
with  $\theta$ a real parameter that can be taken positive without loss of generality. It is convenient to define the creation and annihilation operators
\begin{eqnarray}\nonumber
b &=& \frac{1}{\sqrt{2\theta}} (\hat{x}+i\hat{y}),\\
b^\dagger &=&\frac{1}{\sqrt{2\theta}} (\hat{x}-i\hat{y}),
\end{eqnarray}
that satisfy the Fock algebra $[ b ,b^\dagger ] = 1$. The noncommutative configuration space $\mathcal{H}_c$ is then isomorphic to boson Fock space.

The quantum Hilbert space, in which the physical states of the system are represented, is identified with the set of Hilbert-Schmidt operators acting on noncommutative configuration space
\begin{equation}
\label{qhil}
\mathcal{H}_q = \left\{ \psi(\hat{x},\hat{y}): \psi(\hat{x},\hat{y})\in \mathcal{B}\left(\mathcal{H}_c\right),\;{\rm tr_c}\left[\psi(\hat{x},\hat{y})^\dagger\psi (\hat{x},\hat{y})\right] < \infty \right\}.
\end{equation}
Here ${\rm tr_c}$ denotes the trace over noncommutative configuration space and $\mathcal{B}\left(\mathcal{H}_c\right)$ the set of bounded operators on $\mathcal{H}_c$.  This space has a natural inner product and norm
\begin{equation}\label{inner}
\left(\phi(\hat{x},\hat{y}),\psi(\hat{x},\hat{y})\right) = {\rm tr_c}\left[\phi(\hat{x},\hat{y})^\dagger\psi(\hat{x},\hat{y})\right]
\end{equation}
and forms a Hilbert space \cite{hol}.

To distinguish noncommutative configuration space, which is also a Hilbert space, from the quantum Hilbert space above, we use the notation $|\cdot\rangle$ for elements of noncommutative configuration space, while elements of the quantum Hilbert space are denoted by $\psi(\hat{x},\hat{y})\equiv |\psi)$.  The elements of its dual (linear functionals) are as usual denoted by bras, $(\psi|$, which maps elements of $\mathcal{H}_q$ onto complex numbers by $\left(\phi|\psi\right)=\left(\phi,\psi\right)={\rm tr_c}\left[\phi(\hat{x},\hat{y})^\dagger\psi(\hat{x},\hat{y})\right]$.  We also need to be careful when denoting Hermitian conjugation to distinguish between these two spaces. We reserve the notation $\dagger$ to denote Hermitian conjugation on noncommutative configuration space and the notation $\ddagger$ for Hermitian conjugation on quantum Hilbert space.

The abstract Heisenberg algebra, which reads in two dimensions,
\begin{eqnarray}
\label{heisnc}
\left[{x},{y}\right] &=& i\theta,\nonumber\\
\left[{x},{p}_x\right] = \left[{y},{p}_y\right] &=& i\hbar,\\
\left[{p}_x,{p}_y\right]= \left[x,{p}_y\right] = \left[y,{p}_x\right]&=& 0\nonumber
\end{eqnarray}
is now represented in terms of operators $\hat{X}$,$\hat{Y}$ and $\hat{P}_x$,$\hat{P}_y$ acting on the quantum Hilbert space (\ref{qhil}) with inner product (\ref{inner}), which is the analog of the Schr\"{o}dinger representation of the Heisenberg algebra.  These operators are given by
\begin{eqnarray}
\label{schnc}
\hat{X}\psi(\hat{x},\hat{y}) = \hat{x}\psi(\hat{x},\hat{y})&,& \; \hat{Y}\psi(\hat{x},\hat{y}) = \hat{y}\psi(\hat{x},\hat{y}),\nonumber\\
\hat{P}_x\psi(\hat{x},\hat{y}) = \frac{\hbar}{\theta}[\hat{x},\psi(\hat{x},\hat{y})] &,& \; \hat{P}_y\psi(\hat{x},\hat{y}) = -\frac{\hbar}{\theta}[\hat{y},\psi(\hat{x},\hat{y})],
\end{eqnarray}
i.e., the position acts by left multiplication and the momentum adjointly.  We use capital letters to distinguish operators acting on quantum Hilbert space from those acting on noncommutative configuration space. It is also useful to introduce the following quantum operators
\begin{eqnarray}
\label{qop}
B&=&\frac{1}{\sqrt{2\theta}}\left(\hat{X}+i\hat{Y}\right),\nonumber\\
B^\ddagger&=&\frac{1}{\sqrt{2\theta}}\left(\hat{X}-i\hat{Y}\right),\nonumber\\
\hat{P}&=&\hat{P}_x + i\hat{P}_y,\nonumber\\
\hat{P}^\ddagger &=& \hat{P}_x -i\hat{P}_y.
\end{eqnarray}
We note that $\hat{P}^2=\hat{P}^2_x+\hat{P}^2_y = P^\ddagger P = PP^\ddagger$.  These operators act as follow
\begin{eqnarray}
\label{aqo}
B\psi(\hat{x},\hat{y}) &=& b\psi(\hat{x},\hat{y}),\nonumber\\
B^\ddagger\psi(\hat{x},\hat{y}) &=& b^\dagger\psi(\hat{x},\hat{y}),\nonumber\\
P\psi(\hat{x},\hat{y})&=& -i\hbar \sqrt{\frac{2}{\theta}}[b,\psi(\hat{x},\hat{y})],\nonumber\\
P^\ddagger\psi(\hat{x},\hat{y}) &=& i\hbar\sqrt{\frac{2}{\theta}}[ b^{\dagger},\psi(\hat{x},\hat{y})].
\end{eqnarray}

The interpretation of this quantum system now proceeds as for a standard one.  The only modification required is that position measurement must now be interpreted in the context of a weak measurement (Positive Operator Valued Measure) rather than a strong (Projective Valued Measurement). The essence of the construction is based on the minimal uncertainty states on noncommutative configuration space, which are the normalized coherent states \begin{equation}
\label{cs}
|z\rangle = e^{-z\bar{z}/2}e^{z b^{\dagger}} |0\rangle,
\end{equation}
where $z=\frac{1}{\sqrt{2\theta}}\left(x+iy\right)$ is a dimensionless complex number.  These states provide an overcomplete basis on the noncommutative configuration space.  Corresponding to these states one constructs a state (operator) in quantum Hilbert space as follows
\begin{equation}
\label{braz}
|z )=|z\rangle\langle z|, \quad B|z)=z|z),
\end{equation}
which leads to the natural interpretation of $\left(x,y\right)$ as the dimensionful position coordinates.  These states provide an overcomplete set on the quantum Hilbert space in the form \cite{laure}
\begin{equation}
\label{complstar}
{1}_q=\int \frac{dz d\bar{z}}{\pi} |z)e^{\stackrel{\leftarrow}{\partial_{\bar{z}}}\stackrel{\rightarrow}{\partial_z}}(z|=\int \frac{dx dy}{2\pi\theta} |z)e^{\stackrel{\leftarrow}{\partial_{\bar{z}}}\stackrel{\rightarrow}{\partial_z}}(z|,
\end{equation}
with $\partial_{\bar z}\equiv \frac{\partial}{\partial \bar{z}}$ and $\partial_z\equiv \frac{\partial}{\partial z}$.  This implies that the operators
\begin{equation}
\label{povm}
\pi_z=\frac{1}{2\pi\theta}|z) e^{\stackrel{\leftarrow}{\partial_{\bar{z}}}\stackrel{\rightarrow}{\partial_z}}(z|\;,\quad \int dx dy\;\pi_z=1_q\,
\end{equation}
provide an Operator Valued Measure in the sense of \cite{ber}. We can then give a consistent probability interpretation by assigning the probability of finding the particle at position $\left(x,y\right)$, given that the system is described by the pure state density matrix $\rho=|\psi ) (\psi|$, to be
\begin{eqnarray}
P(x,y)={\rm tr_q}\left(\pi_z\rho\right)=\left(\psi|\pi_z|\psi\right),
\end{eqnarray}
where ${\rm tr_q}$ denotes the trace over quantum Hilbert space. In the light of the developments described below, we shall indeed be able to give a more precise interpretation of this measurement in section \ref{comp}.

The paper is organized as follows: In section \ref{mot} we briefly discuss the path integral formulation of noncommutative quantum mechanics as derived in \cite{sun}. This will allow us to build a picture of the classical mechanics of a noncommutative particle which clearly suggests the notion of extendedness. In section \ref{comp} we present a description of noncommutative quantum mechanics which may be viewed in terms of spatially extended objects. To support this view, we construct the classical action associated with this description in section \ref{class} and show that it coincides precisely with that of \cite{suss} for a free particle. We also discuss the physical interpretation and consequences of this description on the quantum level. Section \ref{concl} summarizes and concludes our discussion.

\section{Hints at extended objects in noncommutative quantum mechanics}
\label{mot}

The first indication of nontrivial structures in noncommutative quantum mechanics comes from the path integral representation of this quantum system.  In \cite{sun} it was shown that the transition amplitude $(z_f, t_f|z_0, t_0)$ can be represented as
\begin{eqnarray}
(z_f, t_f|z_0, t_0)&=&\exp\left(-\vec{\partial}_{z_{f}}\vec{\partial}_{\bar{z}_{0}}\right)
\int \mathcal{D}z\mathcal{D}\bar{z}
\exp({\frac{i}{\hbar}S})
\label{pintegral3}
\end{eqnarray}
where $S$ is the action given by
\begin{eqnarray}
S=\int_{t_{0}}^{t_{f}}dt~ \left[\frac{\theta}{2}\dot{\bar{z}}(t)(\frac{1}{2m}-\frac{i\theta}{2\hbar}
\partial_{t})^{-1}
\dot{z}(t)- V(\bar{z}(t),z(t))\right],
\label{action_ncqm}
\end{eqnarray}
and where we have introduced the dimensionless coordinates $z=\frac{1}{\sqrt{2\theta}}\left(x+iy\right)$.  The potential $V$ is obtained from the expectation value of the normal ordered potential $:V(b^\dagger,b):$ in the coherent state (\ref{cs}) and may thus contain $\theta$-dependent corrections.

Although this representation does not admit a classical limit in the sense of $\hbar\rightarrow 0$, one can still view the classical physics as being determined by the stationary point of the action, while quantum corrections are included through fluctuations around the classical solutions of the equations of motion, as usual. In the case of a quadratic action (free particle and harmonic oscillator) this must indeed yield the exact result, as was demonstrated in \cite{sun}. Motivated by this, let us therefore study the solutions to the equations of motion that follow from the action (\ref{action_ncqm}). From this we should be able to learn more about the nature of the objects we are describing.

Defining the operator $\textbf{K}\equiv\frac{1}{2m}\left(1-i T\partial_{t}\right)$, with
the characteristic time scale $T \equiv m \theta / \hbar$, the equations
of motion read
\be
\frac{\theta}{2} \ddot z (t) = - \textbf{K} V_{\bar z},
\ee
and its complex conjugate. The subscript $\bar z$ denotes the partial derivative. As long as the potential does not depend on time explicitly and has no additional symmetries, at least the total energy is a constant of motion. In this case it is found to be
\be
\label{nlen}
E = m \theta \dot z \dot{\bar z}+V+ i T\int d t\left(\dot z^2 V_{zz}-\text{c.c.}\right).
\ee
One can also construct the total angular momentum,
\be
L = - i m \theta (\bar z \dot z - z \dot{\bar z})+i \int d t\left(z V_{z}-\text{c.c.}\right)+T (V - z V_{z} - \bar z V_{\bar z}),
\ee
but without rotational symmetry it is not a constant of motion. (Note that all integrals are meant as primitives). After rewriting the energy in terms of the dimensionful variables, one recognizes the first two terms as the conventional energy of a point particle, i.e., the sum of kinetic and potential energies. However, there is an additional contribution proportional to $T$. Consequently we note that in this case the sum of the kinetic and potential energies alone is not conserved: an additional contribution is required for the conservation of total energy. This term can only be associated with additional structure.

For rotationally invariant potentials the angular momentum is also a conserved quantity. For such potentials the energy and angular momentum take on a particularly simple and transparent form
\ba
&&E = \frac{1}{2} m \dot\rho^2 + V(\rho) + \frac{\left(L - T V(\rho)\right)^2}{2 m  \rho^2}
,\nonumber\\
&&
L = m  \rho^2 \dot\varphi+T \left(V(\rho) - \rho V'(\rho)\right),
\ea
where we wrote $z$ in terms of a dimensionful polar coordinate $\rho$, $z =\frac{\rho}{\sqrt{2 \theta}} \text{e}^{i \varphi}$, and the prime denotes the ordinary derivative. We recognize the standard kinetic energy, potential energy and angular momentum that one would expect from a point particle described in polar coordinates, but note the presence of noncommutative corrections proportional to $T$, both in the energy and angular momentum. Such corrections can, once again, only be associated with additional structure. Furthermore, in the presence of a potential there is a nontrivial coupling between this structure and the positional degrees of freedom of the object.

These results indicate that structured objects may indeed be the basic entities in noncommutative quantum systems. We proceed by exploring the representation of such quantum systems in a basis that introduces further degrees of freedom (in addition to the positional ones). In the analyses that follows, we demonstrate that one natural way to think of these degrees of freedom is that they relate to the extent of objects in these noncommutative quantum systems.

\section{Noncommutative quantum mechanics viewed in terms of extended objects}
\label{comp}
\subsection{Decomposing the identity; new degrees of freedom}

Let us consider the resolution of the identity on the quantum Hilbert space, $\mathcal{H}_q$: equation (\ref{complstar}). It is possible to decompose the star product $\star:=e^{\stackrel{\leftarrow}{\partial_{\bar{z}}}\stackrel{\rightarrow}{\partial_z}}$ by introducing a further degree of freedom:

\beqa
{1}_q &=& \int \frac{dz d\bar{z}}{\pi} |z)\star(z| \nl
&=& \int \frac{dz d\bar{z}}{\pi} \int \frac{dv d\bar{v}}{\pi} e^{-|v|^2} \ketr{z} e^{\vb\stackrel{\leftarrow}{\partial_{\bar{z}}}+v\stackrel{\rightarrow}{\partial_z}} \brar{z} \nl
&\equiv& \int \frac{dz d\bar{z}}{\pi} \int \frac{dv d\bar{v}}{\pi} \ketr{z,v}\brar{z,v}.
\label{identityzv}
\eeqa

The states $|z,v)$ introduced above have the following additional properties:

\beqa
\ketr{z,v} &=& e^{-v\vb/2} e^{\vb\partial_{\zb}}\ketr{z} \nl
&=& T(z)\ket{0}\bra{v} \nl
&=& e^{\frac{1}{2}(\bar{z}v-\bar{v}z)}\ket{z}\bra{z+v},\,\textnormal{with} \; z,v \in \mathbb{C},
\label{zv}
\eeqa
and
\be
\label{zveigen}
B|z,v)=z|z,v)\quad\forall v.
\ee
Here $T(z)$ is the translation operator defined by

\be
\label{trans}
T\left(z\right)=e^{-\frac{i}{\hbar}(\bar{z}P+zP^\ddagger)},
\ee
with the momentum operators as given in (\ref{aqo}), i.e., the state $|z,v)$ involves the translation of the state $|0,v)\equiv \ket{0}\bra{v}$.

From equation (\ref{zveigen}), we note that the states $|z,v)$ are coherent states in $z$ and as such they are also minimal uncertainty states in configuration space in that they satisfy equality in the $x-y$ uncertainty relation: $\Delta_X\Delta_Y = \frac{\theta}{2}$. This, together with the fact that they are eigenstates of $B$, lends itself to the natural interpretation that these states are the noncommutative analogue of eigenstates of position, since they describe coordinates localized to the minimal uncertainty induced by the commutation relation (\ref{xycommutator}). It is, however, important to note that these statements hold for any $v$ and, indeed, also for any linear combination of these states taken over $v$. It is therefore evident that knowledge of the position of the particle alone, in the sense described in section \ref{intro}, cannot describe a state completely since it provides no information about the right sector. This stands in contrast to a 2 dimensional commutative quantum system in which knowledge of position, i.e., $x$ and $y$, specifies the state completely as these two observables form a maximal commuting set. In the current setting, however, it is clear that knowledge of additional structure, hidden in the right handed sector characterized by $v$, is also required to specify states completely. It is necessary to qualify this statement further: one may forfeit knowledge of the right handed sector by considering the states $|z)=|z\rangle\langle z|$ only, i.e. putting $v=0$, as described in section \ref{intro} and explicitly seen in the resolution of the identity (\ref{identityzv}). The price for this is nonlocality in $z$, in that knowledge of all orders of derivatives of the wavefunction is required for this description. On the other hand, detailed knowledge of the right handed sector allows a manifestly local description in $z$. These two pictures are therefore equivalent: the presence of structure encoded in the right handed sector in the local description, is replaced by the necessity of nonlocal information in the nonlocal description.  The notion of structure is therefore present in both pictures, but the way we access information on that structure differs. One would, however, expect that there is redundancy in the local description and that constraints must arise, since it would be peculiar for the states to be characterized by two independent complex variables. Below we shall see that this is indeed the case.

These considerations become even clearer when one considers position measurements in the weak sense. From (\ref{identityzv}) it is clear that the states $|z,v)$ form an overcomplete set on the quantum Hilbert space. Consequently one may associate a POVM with them, namely

\beq
\pi_{z,v}=\frac{1}{\pi^2}\ketr{z,v}\brar{z,v},
\label{Pizv}
\eeq
as $\pi_{z,v}$ is positive and hermitian on the quantum Hilbert space and integrates to the identity. As a result of this we are able to define the corresponding probability distribution in $z$ and $v$, assuming a pure state density matrix $\rho=\ketr{\psi}\brar{\psi}$,

\be
\label{prob}
P(z,v)={\rm tr}\left(\rho \pi_{z,v}\right)=\brar{\psi}\pi_{z,v}\ketr{\psi}.
\ee

This probability provides information not only about position $z$, but also about a further degree of freedom, $v$, given that the system is prepared in state $\psi$. We could also ask for the probability to find the particle localized at point $z$, without detecting any information regarding $v$. This is simply the sum of the probabilities (\ref{prob}) over all $v$:
\be
\label{probt}
P(z)=\int \frac{dvd\vb}{\pi}\,\, P(z,v)=\brar{\psi}\left[\int \frac{dvd\vb}{\pi}\,\,\pi_{z,v}\right]\ketr{\psi}=\brar{\psi}\pi_{z}\ketr{\psi},
\ee
with
\be
\label{povm1}
\pi_{z}=\int \frac{dvd\vb}{\pi}\,\, \pi_{z,v}=\frac{1}{\pi}\ketr{z}e^{\stackrel{\leftarrow}{\partial_{\bar z}}\stackrel{\rightarrow}{\partial_{z}}}\brar{z},
\ee
and the state $\ketr{z}$ as defined in (\ref{braz}).

Thus our notion of position measurement as set out in section \ref{intro} actually refers to a measurement of position which is insensitive to any other structures that the state may have, while the position measurement encoded in the POVM (\ref{Pizv}) also probes other possible structure of the state and hence provides more detailed information.

Before addressing the specific physical interpretation of the additional degrees of freedom that have been introduced, let us take stock of the discussion in this section. Due to the completeness relation (\ref{complstar}), any state $|\phi)$ can be reconstructed from the overlaps $\left(z|\phi\right)=\left(z,v=0|\phi\right)$ alone. However, as is clear from (\ref{complstar}), this will be a nonlocal construction in $z$ in the sense that it requires knowledge of all the derivatives of the overlap with respect to $z$ at a given point. Since we can construct the state $|\phi)$ itself from knowledge of the overlap $\left(z|\phi\right)$ only, we can certainly also construct the overlap $\left(\chi|\phi\right)$ for an arbitrary state $\chi$ from this information only, albeit in this nonlocal way.  Thus we can access any probability, such as (\ref{prob}), in a representation using the basis $|z)$. Choosing this representation does not, therefore, signal any loss of information: we simply need to access the information through all derivatives of the overlap. On the other hand the completeness relation (\ref{identityzv}) implies that we can also reconstruct any state $|\phi)$ from the overlaps $\left(z,v|\phi\right)$ in a manifestly local way in $z$ that does not require knowledge of higher order derivatives. Subsequently we can also access probabilities such as (\ref{prob}). In the spirit of this discussion, it is important to realize that the two probabilities $P(z,v)$ and $P(z)$ discussed above do not reflect any loss of information, but rather that they address different questions.

It now becomes natural to enquire into the possible physical interpretation of the additional structure uncovered above. As seen in section \ref{mot}, the nonlocal formulation suggests that this new degree of freedom may refer to physical extent, as one would expect the quantum mechanical description of extended objects to require some form of nonlocality. In the local formulation the physical interpretation of this structure is less clear and probably not unique. In order to gain some much-needed intuition regarding the physical interpretation of $v$, we take a look at the classical framework below before returning to the quantum mechanical framework in sections \ref{ang}, \ref{freepart} and \ref{harmos}. In the context of the section to follow, it becomes clear that the classical picture admits a natural interpretation of the variable $v$ as a coordinate that describes deviations of localization from the position $z$, and is intricately connected with the notion of physical extent.

\subsection{The classical picture: insights into the physical interpretation of $v$}
\label{class}

Since the states $|z,v)$ form an overcomplete set of coherent states on the quantum Hilbert space of the noncommutative system, we can derive a path integral action in the standard way \cite{klaud}.  This action is generally given by
\be
\label{locact}
S=\int dt(z,v|i\hbar\partial_t-H|z,v).
\ee
For a noncommutative Hamiltonian of the form $H=\frac{P^2}{2m}+V(X,Y)$ this action can be explicitly computed and is given by
\be
\label{action1}
S=\int dt\left[i\hbar\left(\dot{\bar z} v-\dot z\bar v+\frac{1}{2}\left(\dot{\bar v}v-\dot v\bar v\right)\right)-\frac{\hbar^2}{m\theta}\bar{v}v-V(\bar z,z)\right].
\ee
Let us digress briefly to establish a concrete physical picture. The notion of extended structure can in fact be made explicit in that the action (\ref{action1}) can be precisely identified with that of \cite{suss} in the absence of a potential.  Indeed, consider two particles of mass $m$ and opposite charge $q$ moving in a magnetic field $B$ perpendicular to the plane.  The two particles interact through a harmonic interaction.  Let $z$ be the dimensionful coordinates of one particle and $v$ the dimensionful relative coordinate.  The Lagrangian of this system in the symmetric gauge and in S.I. units is
\be
L=\frac{1}{2}m\dot{\bar z}\dot{z}+\frac{1}{2}m\left(\dot{\bar z}+\dot{\bar v}\right)\left(\dot{z}+\dot{v}\right)+\frac{iqB}{4c}\left(\dot{\bar z}z-\bar z\dot z\right)-\frac{iqB}{4c}\left[\left(\dot{\bar z}+\dot{\bar v}\right)\left(z+v\right)-\left({\bar z}+{\bar v}\right)\left(\dot{z}+\dot{v}\right)\right]-\frac{1}{2}K\bar{v}{v}.
\ee
Introducing the magnetic length $\ell^2=\frac{2\hbar c}{qB}$ and the dimensionless coordinates $\frac{z}{\ell}$ and $\frac{v}{\ell}$ this reduces to
\be
L=\frac{1}{2}m\ell^2\dot{\bar z}\dot{z}+\frac{1}{2}m\ell^2\left(\dot{\bar z}+\dot{\bar v}\right)\left(\dot{z}+\dot{v}\right)+i\hbar\left[\left(\dot{\bar z}v-\bar v\dot z\right)+\frac{1}{2}\left(\dot{\bar v}v-{\bar v}\dot v\right)\right]-\frac{1}{2}K\ell^2\bar{v}{v}.
\ee
In the limit of a strong magnetic field where $\ell\rightarrow 0$ and the kinetic terms may be ignored, this reduces to the Lagrangian (\ref{action1}) upon identifying $K=\frac{2\hbar^2}{m\ell^2\theta}$. In this context $v$ clearly represents the spatial extent of this composite.  Note that for a strong magnetic field the spring tension is very large so that it is difficult to excite the internal modes.  In this case the behaviour is more like that of a stiff rod.

Let us now return to (\ref{action1}) for the case where the potential is nonzero. It is important to note that said potential, which is obtained by writing $X$, $Y$ in terms of $B$ and $B^\ddagger$ and then normal ordering, depends only on $z$, $\bar z$ and not on $v$. Due to the normal ordering this potential will, however, have $\theta$ dependent corrections, i.e., it is not simply the naive potential obtained by replacing the noncommutative variables by commutative ones. In this sense it is different from the classical potential of a point particle to which it reduces in the commutative limit. Upon an appropriate identification of variables, this action reduces to the local action derived in reference \cite{sun} (see eq.(17)).  The properties of this action were already discussed there.  In particular it was found that this is a second class constrained system that yields, upon introduction of Dirac brackets, noncommuting coordinates $z$ and $\bar z$ as one would expect.

The equations of motion are easily obtained as
\ba
\label{eqmot1}
i\hbar \dot{\bar v}-\frac{\partial V}{\partial z}&=&0,\nonumber\\
-i\hbar \dot{v}-\frac{\partial V}{\partial \bar z}&=&0,\nonumber\\
i\hbar \left(\dot{\bar z}+\dot{\bar v}\right)-\frac{\hbar^2}{m\theta}\bar v&=&0,\nonumber\\
-i\hbar \left(\dot{z}+\dot{v}\right)-\frac{\hbar^2}{m\theta}v&=&0.
\ea
Reintroducing the dimensionful variable $z\rightarrow \frac{z}{\sqrt{2\theta}}$, this can be cast in a more recognizable form
\ba
\label{eqmot2}
\ddot z=-\frac{2}{m} \frac{\partial V}{\partial \bar z}-\sqrt{2\theta} \ddot v,\nonumber\\
\ddot{\bar z}=-\frac{2}{m} \frac{\partial V}{\partial z}-\sqrt{2\theta} \ddot{\bar v}.
\ea
(Note that the factor of 2 in the first term is indeed correct as $\partial_z=\frac{1}{2}(\partial_x-i\partial_y)$).  This tells us that, as seen above, the position obeys the standard equations of motion, at least to leading order in $\theta$.  The extra term reflects a coupling of position $z$ to the variable $v$.

It is a simple matter to see that the conserved energy in terms of the dimensionful variable $z$ and dimensionless variable $v$ is given by
\be
\label{en1}
E=\frac{\hbar^2}{m\theta}\bar{v}v+V
\ee
In particular, note from (\ref{locact}) that, as is also reflected in (\ref{en1}), the momentum canonically conjugate to $z$ is $-i\hbar\bar v$.  This reflects a direct relation between $v$, which we associate with spatial extent, and momentum.
Further, we see from (\ref{eqmot1}) that this momentum conjugate to $z$ is not simply $m\dot z$ as for a point particle. This signals that the conserved energy is not just the sum of kinetic and potential energies of a point particle --- a notion that can be made more explicit by rewriting the conserved energy in the following way:
\be
E=\frac{m}{2}\dot{\bar z}\dot{z}+V-m\theta\dot{\bar v}\dot{v}+i\hbar\left(v\dot{\bar v}-\dot{v}\bar{v}\right)
\ee
One can also eliminate $v$ from this, which again yields the nonlocal form (\ref{nlen}).  As expected this consists of the conventional sum of kinetic and potential energies of a point particle with a correction term.  From (\ref{eqmot2}) we see that the dimensionful $z$ has a length scale $\ell_z$, determined by the potential associated with it. Using this in the first two equations of (\ref{eqmot1}), we conclude that the dimensionless $v\sim \frac{\sqrt\theta}{\ell_z}$, which implies the vanishing of the correction terms in the commutative limit. It is, of course, natural that the particular dynamics of a system would govern the positional length scales involved. This generic phenomenon is also demonstrated explicitly on the quantum mechanical framework in section \ref{harmos} in the context of the harmonic oscillator.

Note that for the free particle $v$ and $\bar v$ are simply constants, directly related to the momentum. Thus, for the free particle the spatial extent described by $v$ depends linearly on the momentum, which was also the conclusion reached in \cite{suss}. In section \ref{freepart} we will investigate the quantum mechanical free particle, also in lieu of the connection between momentum and extent. \\

Clearly the arguments above demonstrate that, on the classical level, $v$ describes spatial extent. We shall now proceed by taking this view as a point of departure for the physical interpretation of our noncommutative quantum system. It will be demonstrated that said view is indeed also a natural one on the quantum level.

\subsection{Constraints and differential operators on $\overlapr{z,v}{\psi}$}

As stated earlier, the basis $\ketr{z,v}$ is a suitable one to represent the noncommutative quantum system in terms of overlaps. From (\ref{zv}) it is clear that the bra in $\mathcal{H}_{q}^{*}$ dual to $\ketr{z,v}$ is
\beqa
\brar{z,v} &=& \ket{z+v} \bra{z} e^{\frac{1}{2}(\bar{v}z-\bar{z}v)}  \nonumber \\
&=& e^{-[z\bar{z} + \bar{z}v + \frac{1}{2}v\bar{v}]} e^{(z+v)\bd}\ket{0}\bra{0} e^{\bar{z}b}.
\label{zvdual}
\eeqa

We can now represent the action of the bosonic operators on a state $\ketr{\psi}$ in this basis. For any operator $O$ acting on the quantum Hilbert space, we may introduce left- and right action (denoted by subscripted $L$ and $R$, respectively) as follows:

\be
O_L\psi=O\psi; \;\;O_R\psi=\psi O \;\;\forall \;\psi \in \mathcal{H}_q.
\ee
In this language we have
\beqa
\matrelr{z,v}{\bld}{\psi} & = & e^{\frac{1}{2}(\bar{v}z-\bar{z}v)} \matrel{z}{\bd\psi}{z+v} = \bar{z}\overlapr{z,v}{\psi} \nonumber \\
\matrelr{z,v}{\bl}{\psi}  & = & e^{\frac{1}{2}(\bar{v}z-\bar{z}v)} \matrel{z}{b\psi}{z+v} \,\,= (\diffpa{\bar{z}}+z+v)\overlapr{z,v}{\psi} \nonumber \\
\matrelr{z,v}{\br}{\psi}  & = & e^{\frac{1}{2}(\bar{v}z-\bar{z}v)} \matrel{z}{\psi b}{z+v} \,\,= (z+v)\overlapr{z,v}{\psi}\nonumber \\
\matrelr{z,v}{\brd}{\psi} & = & e^{\frac{1}{2}(\bar{v}z-\bar{z}v)} \matrel{z}{\psi\bd}{z+v} = (\diffpa{v}+\bar{z}+\frac{\bar{v}}{2})\overlapr{z,v}{\psi} = (\diffpa{z}+\bar{z}) \overlapr{z,v}{\psi}
\label{ladders}
\eeqa

Equation (\ref{zvdual}) immediately implies that the overlap $\overlapr{z,v}{\psi}$ must obey the following constraints:
\beqa
&&\left(\diffpa{\bar{v}}+\frac{v}{2}\right )\overlapr{z,v}{\psi}= 0,\label{constr1}\\
&&\left(\diffpa{z}-\diffpa{v}- \frac{\bar{v}}{2}\right) \overlapr{z,v}{\psi}=0.
\label{constr2}
\eeqa
We note that the constraints must hold for \textbf{all} states $\psi$, and thus they restrict which functions $\overlapr{z,v}{\psi}$ are physical in this basis.  We shall refer to this subspace as the physical subspace.  It is important to note, firstly, that these constraints commute with each other and, secondly, that the differential operators associated with the creation and annihilation operators in (\ref{ladders}) all commute with the constraints (\ref{constr1}), (\ref{constr2}), i.e., they leave the physical subspace invariant. This also implies that the differential operator representation in this basis of any operator built from creation and annihilation operators will also leave the physical subspace invariant. This structure allows us to implement the constraints strongly as we shall often do below. With (\ref{ladders}) we have a useful dictionary to represent any operator that is a function of the bosonic operators in terms of derivatives
acting on $\overlapr{z,v}{\psi}$. In the subsequent sections we analyze the angular momentum operator and the hamiltonians of the free particle and the generalized harmonic oscillator by looking at their representations and eigenstates in the basis (\ref{zv}).

\subsection{Angular momentum}
\label{ang}

As an example, let us consider the angular momentum operator that was derived in \cite{laure}. There it
was shown that the generator of rotations is
\beq
L =  X_L P_y - Y_L P_x + \frac{\theta}{2\hbar} PP^\ddagger.
\label{Lz}
\eeq
Noting that
\beq
X_L=\sqrt{2\theta}(\bl+\bld), \quad Y_L=-i \sqrt{2\theta}(\bl-\bld), \quad P_x=\frac{\hbar}{\theta}(Y_L-Y_R), \quad \textnormal{and} \quad P_y=\frac{\hbar}{\theta}(X_R-X_L),
\label{leftrightmomenta}
\eeq
we see that (\ref{Lz}) may also be written as
\beq
L = \hbar\left(\br\brd-\bld\bl\right),
\label{Lzladder}
\eeq
where $\br\brd=(B^\ddagger B)_R$ is the right number operator. We may now write the action of this angular momentum operator in the basis (\ref{zv}) as a differential operator (denoted as $\hat{L}$) by making use of the relevant associations (\ref{ladders}):
\beqa
\hat{L} &=& \hbar \left[(z+v)\left(\diffpa{v}+\bar{z}+\frac{\bar{v}}{2}\right)-\bar{z}\left(\diffpa{\bar{z}}+z+v\right)\right] \nonumber \\
&=& \hbar \left[ v\diffpa{v}+z\diffpa{v}+\frac{|v|^2}{2}+\frac{\bar{v}z}{2}-\bar{z}\diffpa{\bar{z}}\right].
\label{Lzdiff}
\eeqa
Although this representation is unique on the full space, it can be cast in different forms on the physical subspace using the constraints (\ref{constr1}), (\ref{constr2}).  One particular useful and manifestly hermitian form that can be derived is
\beqa
\hat{L} &=& \hbar \left[z\diffpa{z}-\bar{z}\diffpa{\bar{z}}+v\diffpa{v}-\bar{v}\diffpa{\bar{v}} \right]. \nonumber \\
&=& \hat{L}_z + \hat{L}_v,
\label{LzLv}
\eeqa
Here $\hat{L}_z=\hbar\left(z\diffpa{z}-\bar{z}\diffpa{\bar{z}}\right)$ and $\hat{L}_v=\hbar\left(v\diffpa{v}-\bar{v}\diffpa{\bar{v}}\right)$
may be viewed as an orbital angular momentum and an intrinsic angular momentum respectively, since $z$ and $v$ are interpreted, as argued in section \ref{class}, as position and spatial extent of the state, respectively. We thus see the explicit split of total angular momentum into orbital and intrinsic angular momentum, which is clearly in line with the notion of an extended object. It should be emphasized again that (\ref{LzLv}) acts only on the constrained physical subspace. Consequently we should take great care in applying and interpreting this operator. It would be wrong to think that $\hat{L}_z$ and $\hat{L}_v$ are independent operators, i.e. that one could define states on the physical subspace that are simultaneous eigenstates of $\hat{L}_z$ and $\hat{L}_v$. In fact, in lieu of the constraints (\ref{constr1}), (\ref{constr2}) it becomes clear that these two operators do not commute on the physical subspace. Consequently it is not surprising that it is not possible to find physical simultaneous eigenstates of $\hat{L}_z$ and $\hat{L}_v$. To shed some light on this matter, let us consider eigenstates of the total angular momentum operator.

From the form (\ref{Lzladder}) of the angular momentum, it is clear that the states

\beq
\ketr{l} \equiv \sum_{n=0}^{\infty} \alpha_n \ket{n}\bra{n+l}
\eeq

are eigenstates of $L$:

\beqa
L\ketr{l} &=& \hbar\left(\br\brd-\bld\bl\right)\ketr{l} \nonumber \\
&=& \sum_{n=0}^{\infty} \alpha_n \left( \ket{n}\bra{n+l}\bd b - \bd b\ket{n}\bra{n+l} \right) \nonumber \\
&=& \hbar l \ketr{l}.
\eeqa

We now take the overlap of such an angular momentum state with the bra (\ref{zvdual}):

\beqa
\overlapr{z,v}{l} &=& e^{\frac{1}{2}(\bar{v}z-\bar{z}v)} \sum_{n=0}^{\infty} \alpha_n \tr{[\ket{z+v}\bra{z}]\ket{n}\bra{n+l}} \nonumber \\
&=& e^{\frac{1}{2}(\bar{v}z-\bar{z}v)} \sum_{n=0}^{\infty} \alpha_n \overlap{z}{n}\overlap{n+l}{z+v} \nonumber \\
&=& e^{\frac{1}{2}(\bar{v}z-\bar{z}v)} e^{-(|z|^2+|z+v|^2)/2}\sum_{n=0}^{\infty} \alpha_n \frac{\bar{z}^n(z+v)^n}{n!}\frac{(z+v)^l}{\sqrt{(n+l)!/n!}}.
\label{zvl}
\eeqa

This overlap is by construction physical and an eigenfunction of both total angular momentum differential operators (\ref{Lzdiff}) and (\ref{LzLv}). Note, however, that the variables $z$ and $v$ do not decouple in this ``wave function''. We see here explicitly that although (\ref{zvl}) is an eigenstate of total angular momentum, it is not a simultaneous eigenstate of $\hat{L}_z$ and $\hat{L}_v$. Indeed, due to the constraints it is impossible to find such states on the physical subspace of $\mathcal{H}_q$; the requirement of physicality prevents a decoupling of $z$ and $v$ as reflected in (\ref{zvl}).  A physical consequence hereof is that the quantities that we have interpreted as orbital and intrinsic angular momentum, respectively, are not independent. If we consider the clear connection between momentum (motion of the composite) and shape deformation seen in section \ref{class}, this result is reasonable also from a physical point of view. The implication is simply that orbital motion affects shape deformation and consequently intrinsic angular momentum, and vice versa. We infer that $z$ and $v$ cannot be degrees of freedom of a rigid body. This too is in line with the results of \cite{suss}.

\subsection{Free Particle}
\label{freepart}
\def\aa {\sqrt{\frac{\theta}{2\pi\hbar^{2}}}}
\def\ab {\frac{\theta}{4\hbar^{2}}}
\def\ac {\frac{i}{\hbar} \sqrt{\frac{\theta}{2}}}
The hamiltonian of the free particle is simply

\beq
H_{free} = \frac{P^\ddagger P}{2m}= -\frac{\hbar^2}{m\theta}[\bld-\brd]\left[\br-\bl\right].
\label{Hfree}
\eeq

We write the action of (\ref{Hfree}) on a state $\psi$ as a differential operator in the basis
(\ref{zv}) according to (\ref{ladders}):
\beqa
\matrelr{z,v}{H_{free}}{\psi} &=& -\frac{\hbar^2}{m\theta} \left[ \bar{z}- (\diffpa{z}+\bar{z})\right] \left[(z+v)-(\diffpa{\bar{z}}+z+v)\right]
\overlapr{z,v}{\psi} \nonumber \\
&=& -\frac{\hbar^2}{m\theta} \frac{\partial^2}{\partial z\partial \bar{z}}\overlapr{z,v}{\psi}\equiv \hat{H}_{free}\overlapr{z,v}{\psi}
\label{freeHdiff}
\eeqa
Note that the operator $\hat{H}_{free}$ is independent of $v$, which implies a complete decoupling between the structural and positional degrees of freedom for the free particle.  This is not surprising as we know that noncommutativity has no effect on a free particle.

Next we consider eigenstates of the free particle hamiltonian (\ref{Hfree}) as given in \cite{laure},
\beq
\ketr{\psi_{k}} =\aa e^{\ac(\bar{k}b+k\bd)}= \aa e^{-\ab |k|^{2}}e^{\ac k\bd} e^{\ac \bar{k}b},
\label{psik}
\eeq
which can be shown to be eigenstates of the complex momenta $P$ and $P^\ddagger$ with eigenvalues $k$ and $\bar{k}$, respectively. We note
that the overlap of such a momentum state with a basis element (\ref{zv}) is
\beqa
\overlapr{z,v}{\psi_{k}} &=& \aa e^{-\ab |k|^{2}}e^{\frac{1}{2}(\bar{v}z-\bar{z}v)}\bra{z} e^{\ac k\bd} e^{\ac \bar{k}b}\ket{z+v}\nonumber\\
&=& \aa e^{-\ab |k|^{2}} e^{\ac [k\bar{z}+\bar{k}(z+v)]} e^{-\frac{1}{2}|v|^2}.
\eeqa
These states are, by construction, physical and eigenstates of the free particle Hamiltonian (\ref{freeHdiff}).  Note, as expected, the decoupling of the $z$ and $v$ degrees of freedom in this wave function.

Thus, using the POVM (\ref{Pizv}), the probability distribution in $z$ and $v$ for the state $\ketr{\psi_{k}}$ is
\beqa
P(z,v) &=& \matrelr{\psi_{k}}{\Pi_{z,v}}{\psi_{k}}\nonumber \\
&=& \overlapr{\psi_{k}}{z,v} \overlapr{z,v}{\psi_{k}} \nonumber \\
&=& \frac{\theta}{2\pi\hbar^{2}} e^{-\frac{\theta}{2\hbar^{2}}|k|^{2}}e^{\ac [\bar{k}v - \bar{v}k]} e^{-\modsq{v}}.
\label{Pzvfree}
\eeqa

The distribution (\ref{Pzvfree}) has the following properties:
\begin{itemize}
\item{All dependence on $z$ has disappeared. This is to be expected and simply implies that the dynamics of the average position is that of a free particle, as confirmed by (\ref{freeHdiff}).  A measurement of position, which does not enquire about any other possible structure, will therefore yield equal probabilities everywhere.}
\item{The Gaussian $k$-dependence implies a regularization of high momenta, which is in line with the existence of a short length scale $\sqrt\theta$.}
\item{The term $e^{\ac [\bar{k}v - \bar{v}k]}$ represents a momentum-dependent stretching perpendicular to the direction of
motion.  Physically this means that a measurement of the distribution around the center $z$ through the implementation of the POVM (\ref{Pizv}) will yield an asymmetrical momentum dependent distribution, very much as was found in \cite{suss}}
\item{The Gaussian $v$-dependence shows that the spatial distribution around $z$ is confined on the length scale set by $\theta$.  Note that this is a generic feature, which does not depend on dynamics as the Gaussian factor in the wave-function is a consequence of the constraints (\ref{constr1}). We thus expect the local spatial distribution always to be confined to a length scale set by $\theta$, regardless of the particular dynamics, while said dynamics will set the length scale associated with the average position $z$.  We shall indeed see this explicitly for the harmonic oscillator discussed below. In the case of the free particle there is of course no length associated with the average position $z$. One could also view the Gaussian dependence of the wave function on $v$ as arising from harmonic dynamics for $v$ with oscillator length $\sqrt{2\theta}$, which, for small values of $\theta$, corresponds to a very stiff spring constants. It is precisely this picture that emerged from the corresponding classical theory discussed in section \ref{class}}.
\end{itemize}

\subsection{Harmonic oscillator}
\label{harmos}

The harmonic oscillator Hamiltonian discussed in \cite{laure} was
\beq
H = \frac{1}{2m} PP^\ddagger + \frac{1}{2}m\omega^2 (\hat{X}_L^2 +\hat{Y}_L^2).
\label{Hharm1}
\eeq

Note that $\bld \bl = \frac{1}{2\theta}(\hat{X}_L^2 +\hat{Y}_L^2)$. We can thus add a similar term with right action to (\ref{Hharm1}), which yields a slightly more general harmonic oscillator Hamiltonian which we shall consider for the rest of this analysis:

\beqa
H_{h.o.} &=& \frac{1}{2m} PP^\ddagger + m\theta\omega_L^2 (\bld \bl) + m\theta \omega_R^2 (\br \brd) \nonumber \\
&=& \alpha \bld \bl + \beta \brd \br - \gamma (\bld \br + \brd \bl) - m\theta \omega_R^2,
\label{Hharm2}
\eeqa
where

\beq
\alpha = \frac{\hbar^2}{m\theta}+m\theta\omega_L^2, \quad
\beta  = \frac{\hbar^2}{m\theta}+m\theta \omega_R^2, \quad
\gamma = \frac{\hbar^2}{m\theta}.
\label{abg}
\eeq

Note from (\ref{leftrightmomenta}) that the right action term may also be rewritten in terms of momenta and left co-ordinates. This Hamiltonian can therefore also be viewed as a harmonic oscillator with an added magnetic field. Since this Hamiltonian is more general than the one discussed in \cite{laure}, its diagonalization is slightly different to the discussion in \cite{laure} and we briefly digress to describe the diagonalization of this Hamiltonian. To find the eigenstates of (\ref{Hharm2}) we construct a Bogoliubov transformation which introduces new ladder operators of the form
\beq
\left(
  \begin{array}{c}
    A_1 \\
    A_1^\ddagger \\
    A_2 \\
    A_2^\ddagger \\
  \end{array}
\right)
=
\left(
  \begin{array}{cccc}
    \func{cosh}{\phi} & 0 & \func{sinh}{\phi} & 0 \\
    0 & \func{cosh}{\phi} & 0 & \func{sinh}{\phi} \\
    \func{sinh}{\phi} & 0 & \func{cosh}{\phi} & 0 \\
    0 & \func{sinh}{\phi} & 0 & \func{cosh}{\phi} \\
  \end{array}
\right)
\left(
  \begin{array}{c}
    \bl \\
    \bld \\
    \br \\
    \brd \\
  \end{array}
\right)
\label{bogol1}
\eeq
These operators preserve the commutation relations of $B_L$, $B_L^\ddagger$, $B_R$ and $B_R^\ddagger$, i.e.,
$[\bl,\bld]=1;\,[\br,\brd]=-1;\,[\bl,\br]=[\bl,\brd]=0$ and satisfy $[A_1,A_1^\ddagger]=1;\,[A_2,A_2^\ddagger]=-1; [A_1,A_2]=[A_1,A_2^\ddagger]=0$.

Diagonalization of (\ref{Hharm2}) in terms of these new operators fixes the rotation parameter $\phi$ on
\def\rootA{\sqrt {\frac{(\lambda_L+\lambda_R)(4 \hbar^2+ m\theta[\lambda_L+\lambda_R])}{m\theta}}}
\def\rootB{\sqrt {(\omega_L^2+\omega_R^2)[4\hbar^2+m^2\theta^2(\omega_L^2+\omega_R^2)]}}
\def\gamA{1+\frac{m\theta}{2\hbar^2}\left[ \lambda_L+\lambda_R -\rootA  \right]}
\def\gamB{1+\frac{m\theta}{2\hbar^2}\left[ m\theta(\omega_L^2+\omega_R^2) -\rootB  \right]}
\beq
\phi = -\func{arctanh}{\Gamma}, \quad\textnormal{with}\quad
\Gamma=\gamB
\label{phi}
\eeq

Under (\ref{phi}) the inversion of (\ref{bogol1}) and subsequent substitution into (\ref{Hharm2}), we obtain the diagonalized Hamiltonian
\beq
H_{h.o.} = K_1 A_1^\ddagger A_1  + K_2 A_2 A_2^\ddagger + (K_2 - m\theta\omega_R^2)\;
\label{Hharm3}
\eeq
with
\beqa
K_1 &=& \frac{1}{2} \left[ m\theta\omega_L^2-m\theta\omega_R^2+ \rootB \right], \nonumber \\
K_2 &=& \frac{1}{2} \left[ m\theta\omega_R^2-m\theta\omega_L^2+ \rootB \right].
\eeqa

We now have the spectrum for (\ref{Hharm2}):
\beq
E_{n_1,n_2} = n_1 K_1 + (n_2+1) K_2 - m\theta\omega_R^2.
\label{spectrum}
\eeq

Next we construct the vacuum solution, $\ketr{0}$. It is required that
\beqa
A_1\ketr{0} = 0 &\Rightarrow& [ \cosh{(\phi)} \bl + \sinh{(\phi)}\br ] \ketr{0} = 0, \quad \textnormal{and} \nonumber \\
A_2^\ddagger\ketr{0} = 0 &\Rightarrow& [ \sinh{(\phi)} \bld + \cosh{(\phi)}\brd ] \ketr{0} = 0.
\label{annihilate}
\eeqa
Since we know that $\bl\ket{0}\bra{0}=\brd\ket{0}\bra{0}=0$, let us postulate that $\ketr{0}=\mathcal{N}e^{\xi\bld\br}\ket{0}\bra{0}$. Then we have
\beqa
A_1\ketr{0} &=&\mathcal{N}  \left( \cosh{(\phi)}\{[\bl,e^{\xi\bld\br}]+e^{\xi\bld\br}\bl\} +\sinh{(\phi)}\br e^{\xi\bld\br}\right)\ket{0}\bra{0} \nonumber\\
&=& \left( \cosh{(\phi)}\xi\br + \sinh{(\phi)}\br \right)\ketr{0} \quad \textnormal{and}\nonumber \\
A_2^\ddagger\ketr{0} &=&\mathcal{N} \left( \sinh{(\phi)}\bld e^{\xi\bld\br} +\cosh{(\phi)}\{[\brd,e^{\xi\bld\br}]+e^{\xi\bld\br}\brd\} \right) \ket{0}\bra{0} \nonumber \\
&=& \left( \sinh{(\phi)}\bld + \cosh{(\phi)}\xi\bld \right)\ketr{0}
\eeqa
Clearly (\ref{annihilate}) is satisfied if we choose $\xi = -\frac{\sinh{\phi}}{\cosh{\phi}}=-\tanh{\phi}$, i.e. when $\ketr{0}=\mathcal{N}e^{\Gamma\bld\br}\ket{0}\bra{0}$ (see (\ref{phi})). For the normalization of the ground state we note that
\beqa
\overlapr{0}{0} &=& \mathcal{N}^2\, \tr{\left[e^{\Gamma\bld\br}\ket{0}\bra{0}\right]^\ddagger e^{\Gamma\bld\br}\ket{0}\bra{0}} \nonumber \\
&=& \mathcal{N}^2\sum_{n=0}^\infty \sum_{m=0}^\infty \Gamma^{n+m} \tr{\ket{n}\overlap{n}{m}\bra{m}} \nonumber \\
&=& \mathcal{N}^2\sum_{n=0}^\infty \Gamma^{2n} \nonumber \\
&=& \frac{\mathcal{N}^2}{1-\Gamma^2},
\eeqa
where the condition $|\Gamma|<1$ is automatically satisfied due to (\ref{phi}). Thus the correctly normalized ground state is
\beq
\ketr{\psi_0}=\sqrt{1-\Gamma^2}e^{\Gamma\bld\br}\ket{0}\bra{0}.
\label{ground}
\eeq

Finally, excited states can be constructed by applying the appropriate ladder operators from (\ref{bogol1}):
\beq
\ketr{n_1,n_2}_{h.o.}=(A_1^\ddagger)^{n_1}(A_2)^{n_2}\ketr{\psi_0}.
\eeq
In the limit $\omega_R\rightarrow 0$ the above results reduce to those of \cite{laure}.



Next we use (\ref{ladders}) to find the representation of the harmonic oscillator Hamiltonian (\ref{Hharm2}) in the basis (\ref{zv}):

\beqa
\matrelr{z,v}{H_{h.o.}}{\psi} &=& \matrelr{z,v}{\frac{1}{2m} PP^\ddagger + m\theta\omega_L^2 (\bld \bl) + m\theta \omega_R^2 (\br \brd)}{\psi} \nonumber \\
&=& \left[-\frac{\hbar^2}{m\theta} \frac{\partial^2}{\partial z\partial {\bar{z}}}+m\theta\omega_L^2\bar{z}(\diffpa{\bar{z}}+z+v)
+m\theta\omega_R^2(z+v)(\diffpa{v}+\bar{z}+\frac{\bar{v}}{2})\right] \overlapr{z,v}{\psi} \nonumber \\
&:=& \hat{H}_{h.o.} \overlapr{z,v}{\psi}.
\label{Hharmdiff}
\eeqa
We note that the operator $\hat{H}_{h.o.}$ is only hermitian on a function space restricted by the constraints (\ref{constr1}) and (\ref{constr2}), and that its particular form in (\ref{Hharmdiff}) is not unique due to these constraints. Furthermore, it is easy to check that the total angular momentum operator (\ref{LzLv}) commutes with the above Hamiltonian as desired.

As mentioned above, the constraints (\ref{constr1}), (\ref{constr2}) allow the rewriting of (\ref{Hharmdiff}) in many equivalent forms on the physical subspace.  One expects that there should be a manifestly hermitian form, which should reflect the physics more explicitly, and this is indeed the case. Through an appropriate use of constraints the Hamiltonian (\ref{Hharmdiff}) can indeed be rewritten as
\beqa
&&\hat{H}_{h.o.}= -\frac{\hbar^2}{m\theta} \frac{\partial^2}{\partial z\partial {\bar{z}}}+m\theta\omega_L^2\left[|z|^2+\bar{z}(\diffpa{\bar{z}}+\frac{v}{2}-\diffpa{\bar v})+z(-\diffpa{z}+\frac{\bar v}{2}+\diffpa{v})\right]+\nonumber\\
&&m\theta\omega_R^2\left[(z+\frac{v}{2}-\diffpa{\bar v})(\bar{z}+\diffpa{v}+\frac{\bar{v}}{2})\right]\nonumber\\
&&=-\frac{\hbar^2}{m\theta}\frac{\partial^2}{\partial z\partial {\bar{z}}}+m\theta(\omega_L^2+\omega_R^2)|z|^2-m\theta\omega_L^2\left(z\diffpa{z}-\bar{z}\diffpa{\bar z}\right) +m\theta\omega_R^2\left[(\diffpa{ v}+\frac{\bar v}{2})(-\diffpa{\bar v}+\frac{v}{2})-1\right]\nonumber\\
&&+m\theta(\omega_L^2+\omega_R^2)\left(\bar z(-\diffpa{\bar v}+\frac{v}{2})+z(\diffpa{v}+\frac{\bar v}{2})\right).\nonumber\\
\label{Hharmdiff1}
\eeqa

The different contributions in this Hamiltonian have a clear physical meaning: the first two terms represent a normal harmonic oscillator, the only difference being that the frequency is shifted by the right frequency.  If $\omega_R=0$ this is simply the normal harmonic oscillator we would expect.  The third term reflects the standard type of ``Zeeman term'' present in noncommutative systems and is responsible for the well known time reversal symmetry breaking.  The fourth term represents a ``Landau'' Hamiltonian for the variable $v$ with energy scale set by $\omega_R$. The intrinsic dynamics can therefore be viewed as being governed by that of a charged particle moving in a magnetic field with strength set by $\omega_R$.  The last term represents a not unexpected coupling between the variable $v$, describing the spatial extent of the state, and the average position $z$, implying that the spatial extent will be position dependent. A point to note here is that due to the constraint (\ref{constr1}) the wave function $\left(z,v|\psi\right)$ must always contain a Gaussian $e^{-\frac{|v|^2}{2}}$, which implies that this dimensionless parameter is of order $v\sim 1$, i.e., the dimensionful variable $v\sim \sqrt{\theta}$.  Introducing the dimensionful variable $z^\prime=\sqrt{2\theta}z$ one immediately sees from this that the third to last terms are all higher order in $\theta$ and will vanish in the commutative limit to yield the standard commutative harmonic oscillator.

Lastly, let us look at the representation of the ground state (\ref{ground}) in the basis (\ref{zv}):
\beqa
\overlapr{z,v}{\psi_0} &=& \sqrt{1-\Gamma^2}e^{\frac{1}{2}(\bar{z}v-\bar{v}z)}\tr{\ket{z+v}\bra{z} \left[e^{\Gamma\bld\br}\ket{0}\bra{0}\right]} \nonumber\\
&=& \sqrt{1-\Gamma^2}e^{\frac{1}{2}(\bar{z}v-\bar{v}z)} \sum_n^\infty \Gamma^n \overlap{z}{n}\overlap{n}{z+v} \nonumber \\
&=& \sqrt{1-\Gamma^2}e^{\frac{1}{2}(\bar{z}v-\bar{v}z)} e^{\Gamma \bar{z}(z+v)} e^{-\frac{1}{2}(|z|^2+|z+v|^2)}
\eeqa

This implies that the probability distribution in $z$ and $v$ for the ground state is
\beqa
\label{Pzvharm}
P(z,v) &=& \matrelr{\psi_0}{\Pi_{z,v}}{\psi_0}\nonumber \\
&=& \overlapr{\psi_{0}}{z,v} \overlapr{z,v}{\psi_{0}} \nonumber \\
&=& (1-\Gamma^2) e^{\Gamma(2|z|^2+\bar{z}v+\bar{v}z))} e^{-(2|z|^2+|v|^2+\bar{z}v+\bar{v}z)} \nonumber \\
&=& (1-\Gamma^2)\underbrace{e^{-|v|^2}} \underbrace{e^{-2(1-\Gamma)|z|^2}} \underbrace{e^{-(1-\Gamma)(\bar{z}v+\bar{v}z)}} \\
&{}& \qquad\qquad\;\textnormal{(i)}\, \qquad\; \textnormal{(ii)} \qquad\quad \;\;\, \textnormal{(iii)} \nonumber
\eeqa

Let us first investigate this distribution in the standard noncommutative harmonic oscillator limit, i.e. where $\omega_R=0$. For this purpose we define two length-scales,
\beqa
l_\theta &=& \sqrt{2\theta} \nonumber \\
l_{\omega_L} &=& \sqrt{\frac{2\hbar}{m\omega_L}},
\eeqa
where $l_{\omega_L}$ is just the standard harmonic oscillator length scale. Noting that both $z$ and $v$ are dimensionless variables, i.e. $z=\frac{1}{\sqrt{2\theta}}(x+iy)$ (and similarly for $v$), we see that the Gaussian (i) in (\ref{Pzvharm}) decays on a length scale of $l_\theta$. If we associate $v$ with spatial extent this is to be expected as the scale for the spatial extent must be set by the noncommutative parameter. As already remarked this behaviour is quite generic and a consequence of the constraints on the wave function rather than the dynamics. This is also important to ensure that the variable $v$ couples weakly to the variable $z$ for small $\theta$ and decouples in the commutative limit.

From (\ref{phi}) it is clear that
\beq
\Gamma |_{\omega_R=0} = 1+\frac{m\theta}{2\hbar^2}\left[ m\theta\omega_L^2 -\sqrt{\omega_L^2[4\hbar^2+m^2\theta^2\omega_L^2]}  \right],
\eeq
and thus
\beq
1-\Gamma |_{\omega_R=0} = -\left(\frac{l_\theta}{l_{\omega_L}}\right)^2\left[2 \left(\frac{l_\theta}{l_{\omega_L}}\right)^2 - 2\sqrt{1+\left(\frac{l_\theta}{l_{\omega_L}}\right)^4}\right].
\eeq
Since $z = \frac{1}{l_\theta}(x+iy)$, it is clear that under these assumptions the Gaussian (ii) in (\ref{Pzvharm}) decays on a length-scale of $l_{\omega_L}$, with a further dependence on the ratio $\left(\frac{l_\theta}{l_{\omega_L}}\right)$. Comparing this to the case of the commutative harmonic oscillator, where the ground state wave function decays on a length scale of $l_\omega=\sqrt{\frac{2\hbar}{m\omega}}$, this also makes sense: the variable $z$ is associated with the position of a particle moving in a harmonic potential with strength set by $\omega_L$. Finally, we observe that term (iii) in (\ref{Pzvharm}) represents the expected position dependent deformation of the distribution $P(z,v)$.  Note that when $\omega_L=\omega_R=0$ this term vanishes and, as we found for the free particle, there is a decoupling.

In conclusion, the representation of the ground state (\ref{ground}) for the case $\omega_R=0$ in the basis (\ref{zv}) shows explicitly that there are two length scales involved in the problem: the fundamental harmonic oscillator length scale \textit{as well as} the length scale set by the noncommutative parameter $\theta$. As discussed in \ref{class}, it is of course generic that the particular dynamics sets the positional length scale of a problem. In the case where $\omega_R\neq0$, the decay of the Gaussian term (ii) in (\ref{Pzvharm}) would be governed by two length scales: $l_{\omega_L}$ and $l_{\omega_R}$.

\section{Discussion and conclusions}
\label{concl}

By considering the path integral representation found in \cite{sun} and looking at constants of motion, we showed that already in a nonlocal, unconstrained description of noncommutative quantum mechanics there are hints at extended objects in the theory. It was then demonstrated explicitly in the classical picture that the energy contains correction terms proportional to the noncommutative parameter.  These corrections could also be cast in a local or nonlocal form. In the local formulation the Lagrangian of a free particle coincides precisely with that of two oppositely charged particles interacting with a harmonic potential and moving in a strong magnetic field. Using these results as a primer, we proceeded to show that an interpretation of noncommutative quantum mechanics in terms of extended objects with additional structure is indeed a natural one. Two equivalent descriptions of noncommutative quantum mechanics exhibiting this feature, namely a local and the aforementioned nonlocal one, were found. In the local description the additional structure of the particle was made explicit through the introduction of an additional degree of freedom, while it was encoded in higher order derivatives in the nonlocal description.


\section{Acknowledgements}

This work was supported under a grant of the National Research Foundation of South Africa. We also acknowledge useful discussions with J Govaerts, JN Kriel, S Vaidya and B Chakraborty.  FGS also acknowledges the warm hospitality of the Centre for Particle Physics and Phenomenology of the Catholic University of Louvain and the Centre for High Energy Physics of the Indian Institute of Science where parts of this work were completed.


\end{document}